\documentclass[10pt]{article}

\usepackage[letterpaper,textwidth=5.5in,textheight=9in,left=1.5in,top=1in,
            headheight=12pt,headsep=25pt,footskip=30pt]{geometry}
\usepackage[T1]{fontenc}
\usepackage{amsmath}
\usepackage{newtxtext}
\usepackage{newtxmath}
\usepackage{microtype}
\usepackage{graphicx}
\graphicspath{{figures/}}
\usepackage{booktabs}
\usepackage{array}
\usepackage{textcomp}
\usepackage{url}
\usepackage[authoryear,round]{natbib}
\usepackage[font=small,labelfont=bf]{caption}
\usepackage{titlesec}
\usepackage{xcolor}
\usepackage[colorlinks=true,linkcolor=black,citecolor=black,urlcolor=black]{hyperref}

\titleformat{\section}{\normalfont\large\bfseries}{\thesection}{1em}{}
\titlespacing*{\section}{0pt}{2.5ex plus 1ex minus .2ex}{1.5ex plus .3ex}
\titleformat{\subsection}{\normalfont\normalsize\bfseries}{\thesubsection}{1em}{}
\titlespacing*{\subsection}{0pt}{2ex plus 1ex minus .2ex}{1ex plus .2ex}
\titleformat{\paragraph}[runin]{\normalfont\normalsize\bfseries}{\theparagraph}{1em}{}
\titlespacing*{\paragraph}{0pt}{1ex plus .5ex}{1em}
\newcommand{\mmd}{mm\,d$^{-1}$}
\newcommand{\FSS}{$\mathrm{FSS}_{20}$}
\newcommand{\CSI}[1]{$\mathrm{CSI}_{#1}$}

\makeatletter
\renewcommand{\maketitle}{%
  \par\begingroup\centering
  \vskip 0.1in\hrule height 4pt\vskip 0.25in
  {\LARGE\bfseries\@title\par}
  \vskip 0.29in\hrule height 1pt\vskip 0.3in
  {\normalsize\@author\par}
  \vskip 0.3in\endgroup
  {\let\thefootnote\relax\footnotetext{Preprint.}}}
\makeatother
\renewenvironment{abstract}{%
  \begin{center}{\large\bfseries Abstract}\end{center}\vspace{-1.5ex}%
  \begin{list}{}{\setlength{\leftmargin}{0.5in}\setlength{\rightmargin}{0.5in}%
                 \setlength{\parsep}{4pt}\setlength{\topsep}{0pt}}\item\relax}%
  {\end{list}\vspace{2ex}}

\title{Statistical Downscaling of Daily Precipitation from ERA5 Atmospheric Fields to IMERG over West Africa}
\author{%
  {\bfseries David Sagna}$^{1}$\qquad{\bfseries Redouane Lguensat}$^{2}$\qquad{\bfseries Samo Diatta}$^{1}$\\[1ex]
  $^{1}$Laboratoire d'Oc\'eanographie et des Sciences de l'Environnement et du Climat (LOSEC),\\
  Universit\'e Assane Seck de Ziguinchor, BP~523, Ziguinchor, Senegal\\[0.5ex]
  $^{2}$Institut Pierre-Simon Laplace (IPSL), Sorbonne Universit\'e, IRD, 75005 Paris, France\\[1ex]
  Correspondence: \texttt{d.sagna5060@zig.univ.sn}}

\begin{document}

\maketitle

\begin{abstract}
Global reanalyses are too coarse to resolve the convective systems that produce most West African monsoon rainfall. In this work, we use U-Net based architectures to downscale daily ERA5 fields (0.25°) to IMERG V07 satellite precipitation (0.1°) over West Africa for the June to October seasons of 2000 to 2024, using geopotential, temperature, humidity and wind at three pressure levels together with elevation and land fraction. Adding ERA5 precipitation as input slightly improves performance for this particular challenging region. We find that all the deterministic networks investigated for this work are too dry and lose almost all observed variance below a few hundred kilometers, which motivated our use of a GAN-based alternative. We also investigated permutation importance for the U-Net that ranks 500 hPa specific humidity first, consistent with the sensitivity of deep convection to mid-tropospheric moisture.
\end{abstract}

\section{Introduction}\label{sec:intro}

Rainfall in West Africa is concentrated in the June--October monsoon season and
varies strongly from year to year and from one region to the next, which makes
it a first-order constraint on agriculture, water resources and flood
risk~\citep{nicholson2013,sultan2013,sylla2016,polasky2023}. Studies of these
impacts need daily precipitation fields fine enough to resolve individual
convective systems. Global climate models and reanalyses provide the large-scale
circulation at resolutions of tens to hundreds of kilometres, but the
precipitation of climate models is biased, notably because deep convection is
parameterised~\citep{kendon2019}. Regional climate models reduce the resolution
gap at a large computational cost~\citep{giorgi2019,gutowski2016,prein2015}.
Statistical downscaling instead learns an empirical relation between large-scale
predictors and local precipitation~\citep{fowler2007,maraun2018}. In the
perfect-prognosis formulation, this relation is learnt from observed or
reanalysed predictors and observed precipitation, and no simulated precipitation
enters the input, so that the trained model can later be applied to the
circulation of a climate model without inheriting the biases of its
rainfall~\citep{maraun2018}.

Convolutional neural networks have become a common tool for this task~\citep{rampal2024review}.
\citet{banomedina2020} compared network configurations for perfect-prognosis
downscaling over Europe, and the resulting DeepESD model was used to downscale
an ensemble of climate projections~\citep{banomedina2022}. Emulators of regional
climate models built on U-Nets~\citep{ronneberger2015} learn the high-resolution
fields of a regional model from its large-scale predictors~\citep{doury2023},
including daily precipitation~\citep{doury2024}. \citet{doury2024} also proposed
an asymmetric loss that penalises the underestimation of rainfall that is heavy
for its location, which \citet{gonzalezabad2025} then used to compare deep
downscaling models. Other studies refine a coarse precipitation field instead,
in a super-resolution formulation~\citep{vandal2017,harder2025}, add attention
mechanisms to the network~\citep{sun2020}, or train generative models to restore
the fine-scale variability that deterministic networks smooth
away~\citep{leinonen2021,harris2022,rampal2025james,mardani2025,lee2026}.
\citet{glawion2025}, for example, downscaled ERA5 precipitation over the globe
to 2\,km and 10\,min with a generative network trained on radar observations
over Germany and evaluated over the United States and Australia.

Over West Africa, statistical downscaling has mostly been applied to station
data and to climate projections~\citep{sarr2015,polasky2023}. \citet{ebiendele2025}
trained U-Nets in a perfect-prognosis framework, from ERA5 predictors upscaled to
200\,km to the 25\,km precipitation of the Climate Hazards Group InfraRed
Precipitation with Station data (CHIRPS), and applied them to the output of a
global climate model. Over sub-Saharan Africa, \citet{lee2026} used a diffusion
model to recover 0.05\textdegree{} CHIRPS precipitation from its own
coarse-grained 0.25\textdegree{} version and ERA5 predictors, and found that
deterministic networks trained with a squared error smooth the fields and
suppress the extremes. \citet{harder2025} downscaled hourly ERA5 fields,
precipitation among them, to the Integrated Multi-satellitE Retrievals for GPM
(IMERG~V07) at 0.1\textdegree{} and tested their models on West Africa, among
other regions. Each of these studies differs from ours in an important respect. \citet{lee2026}
recover a field from a coarse-grained copy of itself, which already places the
rain, whereas the observed rain never enters our inputs, even coarse-grained: our
models start from the atmospheric state given by the geopotential, temperature,
humidity and wind fields of the ERA5 reanalysis~\citep{hersbach2020}, alone, as
in the perfect-prognosis formulation, or with ERA5 precipitation, the reanalysis
model's own estimate of the rain, whose errors of position and intensity they
must then correct. \citet{ebiendele2025} upscaled the predictors to 200\,km to
predict CHIRPS at 25\,km, whereas ours keep the 0.25\textdegree{} resolution of
ERA5 to predict daily IMERG~V07 precipitation~\citep{imergv07} at its native
0.1\textdegree{}, about 11\,km. \citet{harder2025}, whose inputs and target are
the closest to ours, benchmarked the transfer of models between regions, whereas
our study is built around the West African monsoon: the models are trained on the
region itself, at the daily scale and for the June--October season, and
evaluated zone by zone, from the humid Guinean coast to the arid Sahel, and month
by month, against regional baselines, with an analysis of the predictors behind
their skill. At 0.1\textdegree{}, much of the day-to-day rainfall variability
comes from mesoscale convective systems, which produce most of the rainfall of
the Sahel~\citep{mathon2002}. The large-scale fields may constrain their exact
position only weakly, so that the attainable skill is limited and the choice of
baselines and of the evaluation protocol matters as much as the choice of
network.

We therefore ask five questions. First, how much skill do U-Nets add over
simpler baselines, namely climatology, ERA5 precipitation interpolated to
0.1\textdegree{} or downscaled by bias correction and spatial disaggregation, and
DeepESD, which uses the predictors of the U-Nets averaged to 1\textdegree?
Second, does adding ERA5 precipitation as an input improve the U-Net for this region? Third,
what does quantile mapping add, when its design is chosen without using
the test period? Fourth, can a generative adversarial network (GAN) trained on
top of a U-Net restore the fine-scale variability of the rainfall fields without
losing skill, and give a calibrated ensemble? Fifth, which
predictors does the network rely on? All models share the same data and
evaluation protocol: all choices are based on two validation seasons, and four
test seasons are kept for the evaluation (Section~\ref{sec:eval}).

\section{Data}\label{sec:data}

\subsection{Study domain and period}

\begin{figure}[t]
\centering
\includegraphics[width=0.85\textwidth]{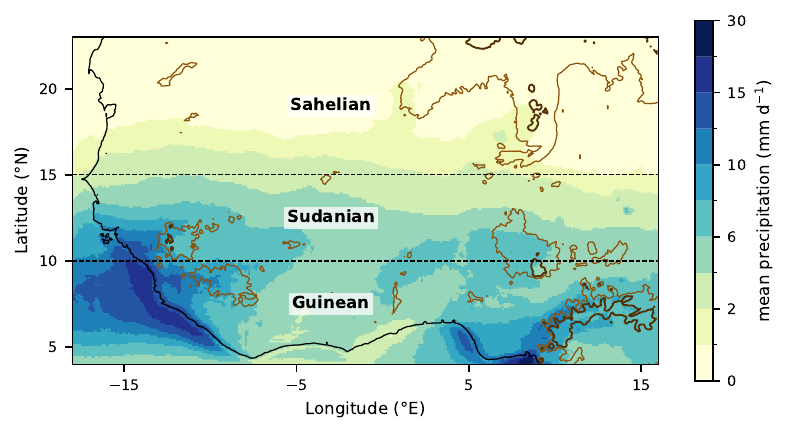}
\caption{Study domain and mean IMERG~V07 precipitation over the June--October
seasons of 2000--2024 (\mmd). The black line is the coastline (land fraction
0.5), the brown contours mark elevations of 500 and 1000\,m, and the dashed
lines at 10\textdegree N and 15\textdegree N separate the Guinean, Sudanian
and Sahelian zones.}\label{fig:domain}
\end{figure}

The study domain covers 4--23\textdegree N and 18\textdegree W--16\textdegree E
(Figure~\ref{fig:domain}). We divide it into three latitude bands, named after
the climatic zones they approximately span: a Guinean zone (4--10\textdegree N),
a Sudanian zone (10--15\textdegree N) and a Sahelian zone (15--23\textdegree N),
whose northern part lies in the Sahara. The zones include the ocean pixels of
each band. Mean June--October precipitation exceeds 15\,\mmd{} along the coast
of Guinea and Sierra Leone and around Mount Cameroon, and decreases northwards
to less than 1\,\mmd{} in the Sahara.

The period covers the June--October seasons of 2000 to 2024 (3\,825 days). The
season of 2000 is the first in the archive we used, and that of 2024 is the last complete season at the time of
this study. The seasons are split chronologically into a training period
(2000--2018, 2\,907 days), a validation period (2019--2020, 306 days) and a test
period (2021--2024, 612 days).

\subsection{Target: IMERG V07}

The target is the IMERG~V07B Final Run half-hourly precipitation of the Global
Precipitation Measurement mission~\citep{imergv07}, at its native
0.1\textdegree{} resolution. Daily precipitation is 24 times the mean of the
48 half-hourly rates of the UTC day; all 48 files were available on every day
of the period. The study domain contains $190\times340$ IMERG cells.

\subsection{Predictors: ERA5}

The predictors come from the hourly ERA5 analyses on pressure
levels~\citep{hersbach2020,era5pl} at 0.25\textdegree: geopotential ($Z$),
temperature ($T$), specific humidity ($Q$), relative humidity ($R$) and the
zonal ($U$) and meridional ($V$) wind at 850, 700 and 500\,hPa, i.e.\ 18
channels. Daily values are means over the same UTC day as the IMERG target,
computed with the trapezoidal rule from the 25 hourly analyses from 00 to
24\,UTC. Relative humidity is missing for part of the period in the archive we
used. To keep the series homogeneous, we diagnose it for all years from the
hourly temperature and specific humidity, with the saturation vapour pressure of
the ECMWF model, a temperature-weighted combination of the values over water and
ice~\citep{ecmwf2016ifs}, and then average it. ERA5 precipitation, used as a
baseline and as an optional input, is the sum of the hourly forecast
accumulations valid over the same UTC day~\citep{era5sl}.

The U-Nets also receive two static fields, mean land elevation and land fraction
on the 0.1\textdegree{} grid, aggregated from the 60~arc-second version of the
ETOPO~2022 global relief model~\citep{etopo2022}.

\subsection{Grids and normalisation}

The U-Nets work on a $192\times352$ grid at 0.1\textdegree, which covers the
study domain with a margin of one to six cells, so that both dimensions are
divisible by 16. Losses and scores are computed on the study domain only. The
ERA5 fields are kept at 0.25\textdegree{} and interpolated bilinearly to the
0.1\textdegree{} grid inside the data pipeline, with weights computed from the
geographic coordinates of both grids. Each ERA5 channel is standardised with the
mean and standard deviation of the training period, and the two static fields
with their spatial mean and standard deviation.

\section{Methods}\label{sec:methods}

\subsection{U-Net models}

The U-Net~\citep{ronneberger2015} has four down-sampling steps, with 64, 128,
256 and 512 channels in the encoder and 1024 in the bottleneck. Each block
applies two $3\times3$ convolutions with instance
normalisation~\citep{ulyanov2016} and leaky ReLU activations, and spatial
dropout (rate 0.1) between them. The decoder up-samples bilinearly, halves the
number of channels with a $1\times1$ convolution and concatenates the skip
connection. A final $1\times1$ convolution followed by a ReLU gives daily
precipitation in \mmd. The network has 29.0\,M parameters and follows the
perfect-prognosis formulation. A second configuration, the U-Net with ERA5
precipitation, takes $\ln(1+p)$, where $p$ is ERA5 precipitation, as a 19th
dynamic input; it leaves the perfect-prognosis formulation and measures what the
coarse precipitation field adds. With or without ERA5 precipitation, adding
convolutional block attention modules~\citep{woo2018} after every block and
attention gates~\citep{oktay2018} on the skip connections lowered the correlation
on the validation seasons by 0.005 and the fractions skill score at 20\,\mmd{}
(Section~\ref{sec:eval}) by 0.011 to 0.015 (means over five training runs), so
we did not keep them.

\subsection{Loss function}\label{sec:loss}

All deterministic networks are trained with the asymmetric loss of
\citet{doury2024},
\begin{equation}\label{eq:asym}
\mathcal{L} = \frac{1}{N}\sum_{i=1}^{N}\Big[\,|y_i-\hat y_i|
  + G_i(y_i)^2\,\max(0,\,y_i-\hat y_i)\Big],
\end{equation}
where the sum runs over the $N$ pixel-days of a batch in the study domain,
$y_i$ and $\hat y_i$ are the observed and predicted precipitation at
pixel-day~$i$, and $G_i$ is the cumulative distribution function of a gamma
distribution fitted to the observed wet days ($y\geq1$\,\mmd) of the training
period at the pixel of~$i$. The second term penalises underestimation,
increasingly so as the observed value becomes more extreme for its location. The
shape parameter is estimated with the closed-form approximation
$k=\big(3-s+\sqrt{(s-3)^2+24s}\,\big)/(12s)$ of its maximum-likelihood value,
with $s=\ln\overline{y}-\overline{\ln y}$ over the wet days, and the scale as
$\overline{y}/k$; the pixels with fewer than 20 wet days share a single fit.
These details of the fit differ from those of \citet{doury2024}. The first term
is an absolute error, whose minimiser is the conditional median, so the loss
pulls predictions towards a central value of the conditional distribution and
raises them only for heavy rain; we return to this point in
Section~\ref{sec:disc-loss}.

\subsection{Training}\label{sec:training}

We train the U-Nets with AdamW~\citep{loshchilov2019}, a learning rate of
$3\times10^{-4}$ and a weight decay of $10^{-4}$, batches of 8 days, gradient
clipping at norm~1 and a cosine decay of the learning rate~\citep{loshchilov2017}
to $10^{-6}$ over 40 epochs, and keep the epoch with the lowest validation loss
(Appendix~\ref{app:training}, Figure~\ref{fig:training}). A training run
takes about 16 minutes on one NVIDIA RTX~5090 GPU (Appendix~\ref{app:timing}).

\subsection{Residual Wasserstein GAN}\label{sec:gan}

Generative adversarial networks have been used to generate the fine-scale
variability of precipitation fields~\citep{leinonen2021,harris2022}.
\citet{rampal2025james} proposed a residual GAN with an intensity constraint
that makes its skill robust to the weight of the adversarial loss, and
\citet{rampal2025intercomparison} refined it (their ResGAN-v2). Our residual
Wasserstein GAN (residual WGAN, or simply the GAN below) uses the loss and the
optimisation of ResGAN-v2 on top of the U-Net with ERA5 precipitation, with the
differences listed at the end of this section. The trained U-Net is frozen and
provides a deterministic prediction~$\hat y$. A smaller U-Net generator (widths
32 to 512, 7.2\,M parameters, without dropout) receives the inputs of the U-Net,
ERA5 precipitation included, and $\ln(1+\hat y)$; eight channels of Gaussian
noise are concatenated to its features before its first two up-sampling steps,
on the $12\times22$ and $24\times44$ grids. It predicts a residual $\hat r$ in
$\ln(1+y)$ space, and the generated field is
$\exp\!\big(\ln(1+\hat y)+\hat r\big)-1$, set to zero where negative.

A conditional PatchGAN critic~\citep{isola2017}, with 2.8\,M parameters and a
receptive field of 70 cells, scores the residual of one member together with the
inputs.
It is trained with the Wasserstein loss and a gradient penalty of
weight~10~\citep{arjovsky2017,gulrajani2017}, with three critic steps per
generator step. The generator loss is the mean squared error between the mean of
eight generated residuals and the observed residual, with land pixels weighted
eight times more than ocean pixels, plus 0.01 times the adversarial loss, plus
the squared error between the domain maxima of $\ln(1+y)$ in each generated
member and in the observation (the intensity constraint of
\citealt{rampal2025james}). Taking the error on the mean of several members, as
\citet{harris2022} and \citet{rampal2025intercomparison} did, rather than on
each member, divides the penalty that the squared error puts on differences between the
members by their number, but does not reward such differences. Both networks are
trained with Adam~\citep{kingma2015}, a learning rate of $1.5\times10^{-4}$,
$\beta_1=0.5$ and $\beta_2=0.9$, on batches of 8 days, with the learning rate of
the generator multiplied by 0.995 every 1000 steps, for 200 epochs (24\,200
generator steps), and the last epoch is kept. The GAN is trained once, on the
trained U-Net with ERA5 precipitation. Our implementation differs from
ResGAN-v2 in its critic, a PatchGAN rather than a network with dense layers over
the whole field; in its generator, a U-Net with instance normalisation, without
the day-of-year conditioning and the attention of ResGAN-v2; in keeping the
U-Net frozen; in its predictors and training data (Section~\ref{sec:data}),
2\,907 days against the decades of regional-model output used by
\citet{rampal2025intercomparison}; and in its smaller number of generator steps.
The GAN generates ten members for every day; the first member is scored like the
deterministic models, and the ten members are evaluated as an ensemble.

\subsection{Baselines}\label{sec:baselines}

\paragraph{Climatology.} The prediction is the mean precipitation of each
calendar month at each pixel over the training period.

\paragraph{ERA5 precipitation.} The prediction is the ERA5 daily precipitation,
interpolated bilinearly to the 0.1\textdegree{} grid. It uses no training data
and measures the skill that the reanalysis already contains.

\paragraph{BCSD.} Bias correction and spatial
disaggregation~\citep[BCSD;][]{wood2002,wood2004} is a standard statistical
downscaling of the precipitation of a model, which we apply to ERA5
precipitation in its daily form~\citep{thrasher2012,thrasher2022}. IMERG is first
averaged over the 0.25\textdegree{} cells of ERA5, each 0.1\textdegree{} cell
weighted by its overlap with the cell. For each 0.25\textdegree{} cell and each
day of the season, ERA5 precipitation is then mapped onto these averages by
empirical quantile mapping, fitted on the days of the training seasons within
15~days of the same date. Dry days of ERA5 are ranked at random among
themselves, and values above the largest one of this sample are set to the
largest observed value, which did better on the validation seasons than
extrapolation with a constant ratio or a Gumbel law. Finally, the corrected
value is divided by the mean of the observed averages over the same window of
training days (at least 0.1\,\mmd), and this ratio is interpolated bilinearly
to the 0.1\textdegree{} grid and multiplied by the mean IMERG precipitation of
each pixel over the same window. BCSD thus combines the corrected ERA5 field with the observed
fine-scale climatology; like the ERA5 precipitation baseline, it does not use the
atmospheric predictors.

\paragraph{DeepESD.} DeepESD is the model of \citet{banomedina2022} in its
precipitation configuration, as implemented by its authors in the
\texttt{deep4downscaling}
library\footnote{\url{https://github.com/SantanderMetGroup/deep4downscaling}}:
three $3\times3$ convolutions with 50, 25 and 1 filters and ReLU activations,
followed by a linear layer that predicts every target pixel. The original model
predicts the parameters of a Bernoulli--gamma distribution and is trained by
maximum likelihood; like \citet{gonzalezabad2025}, we instead predict the amount
directly, train the network with the asymmetric loss and set negative
predictions to zero. A dense layer from the 0.25\textdegree{} fields to 64\,600
pixels would be very large, so the 18 predictors are averaged over
1\textdegree{}~$\times$~1\textdegree{} boxes ($19\times34$) and standardised at
each grid point over the training period; unlike the implementation of its
authors, the convolutions have no padding. The model has 23.6\,M parameters. It
is trained with Adam~\citep{kingma2015}, a learning rate of $10^{-4}$, batches of
64 days and early stopping with a patience of 60 epochs.

\subsection{Quantile mapping}\label{sec:qm}

We also evaluate every model except climatology after an empirical quantile
mapping~\citep{gudmundsson2012}. We fit the mapping on the predictions of each
model for the validation seasons, which were used in training only to select the
epoch of the networks and the treatment of the upper tail in BCSD, and apply
it to the test seasons. For each pixel and calendar month, we estimate the
predicted and observed quantiles at levels $0,0.01,\dots,0.99$ from a sample
that pools the $5\times5$ neighbouring pixels and the previous and next months,
i.e.\ 3\,000 to 4\,600 values. We map a prediction by linear interpolation
between the predicted and observed quantiles and set values above the predicted
99th percentile to the observed 99th percentile.

We chose this design by a two-fold cross-validation within the validation
seasons (fitting on 2019 and applying to 2020, and conversely), among variants
that differ in the treatment of the upper tail and in the pooling
(Appendix~\ref{app:qm}, Table~\ref{tab:qmcv}). In particular, extending the correction above the
highest fitted quantile with a constant ratio did worse than the retained design
on every score of this cross-validation (Section~\ref{sec:disc-loss}).

For the ERA5 precipitation baseline, which has no training period, fitting the
mapping on all the seasons of 2000--2020 instead changes its fractions skill
score (\FSS{}) and critical success indices (\CSI{20} and \CSI{40}) on the test
seasons by at most 0.005 (\FSS{} 0.446 instead of 0.441, \CSI{40} 0.140 instead
of 0.139). Its comparison with the trained models therefore does not depend on
the short calibration period.

\subsection{Evaluation}\label{sec:eval}

All choices of models, hyperparameters and post-processing were based on the
validation seasons, and the test seasons were used only for the evaluation.
Scores are computed over all pixel-days of the study domain in the test seasons
(612 days) and, separately, over land pixels (land fraction at least 0.5), over
the three zones of Figure~\ref{fig:domain} (land and ocean) and for each month.

We report the root-mean-square error (RMSE), the mean absolute error (MAE), the
mean bias (prediction
minus observation), the Pearson
correlation $r$ over all pixel-days and the ratio of the predicted to the
observed 99th percentile of the pixel-day values (P99 ratio). Because the pooled
correlation includes the spatial pattern of mean rainfall, which climatology
alone captures, we also give the mean over pixels of the correlation between the
predicted and observed daily series. The event scores are the fractions skill
score~\citep{roberts2008} at 20\,\mmd{} over $5\times5$-pixel neighbourhoods
(\FSS) in its aggregated form, $1-\sum(P-O)^2/\sum(P^2+O^2)$, where $P$ and $O$
are the predicted and observed fractions and the sums run over all pixels and
days, with the rain outside the zone set to zero when a zone is scored. The
critical success index, $h/(h+m+f)$ for $h$ hits, $m$ misses and $f$ false
alarms, is given at 5, 20 and 40\,\mmd{} (\CSI{5}, \CSI{20}, \CSI{40}), and at
40\,\mmd{} we also give the probability of detection, $h/(h+m)$ (POD), and the
false alarm ratio, $f/(h+f)$ (FAR)~\citep{wilks2019}.

We compare the spatial structure of the fields through their radially averaged
power spectral density (RAPSD), computed for each day with the \texttt{rapsd}
function of the pysteps library~\citep{pulkkinen2019}. This function averages
the power over rings of equal wavenumber index, which matches a physical
wavelength only for a square field, so we split the domain into two overlapping
windows of $190\times190$ cells (west and east) and average the spectra over
both windows and all test days. The fields are not periodic, with rain along the
southern edge of each window and little along its northern edge. We therefore
remove the mean of each window and multiply it by a two-dimensional Hann window
before the transform; without this taper, the jump between the edges leaks power
to all wavenumbers and dominates the small scales of smooth fields. We compute
wavelengths with a nominal spacing of 11.1\,km per 0.1\textdegree{} cell, which
ignores that the zonal spacing is up to 8\% shorter at 23\textdegree N.

The ensemble of the GAN is evaluated with the continuous ranked probability
score (CRPS) of its ten members, in the fair form of \citet{ferro2014}, which reduces to the
MAE for a deterministic model~\citep{gneiting2007}. It is also evaluated with
the spread-skill ratio (SSR), the ratio of the ensemble spread (the square root
of the mean over pixel-days of the ensemble variance, computed with $M-1$ in its
denominator and multiplied by $(M+1)/M$ for $M$ members) to the RMSE of the
ensemble mean, which is about 1 for a calibrated
ensemble~\citep{fortin2014}, and with rank histograms~\citep{hamill2001}, in
which amounts below 0.1\,\mmd{} count as zero and ties, mostly dry pixels, are
broken at random. The histograms are computed over all pixel-days and over the
pixel-days on which IMERG or at least one member has at least 1\,\mmd; this
condition, unlike a condition on the observation alone, treats the observation
and the members alike, so that the histogram of a calibrated ensemble stays
flat. To see at which scales the ensemble lacks spread, we compare the power
spectra of the difference between two members and of the difference between a
member and IMERG: if the observation behaves like one more member, as it should
for a calibrated ensemble, the two spectra coincide at every scale. We also
compute the spread-skill ratio of fields averaged over boxes of up to
2\textdegree{} and, as a measure of the structure of the fields, the number and
the size of the connected areas with at least 20\,\mmd{} on each day.

Predictor importance is estimated for the U-Net by
permutation~\citep{breiman2001}: the field of one input channel is replaced by
the field of a randomly chosen other test day, the increase in RMSE is recorded,
and the procedure is repeated three times for each of the 18 dynamic channels.

\section{Results}\label{sec:results}

\subsection{Overall skill}

Table~\ref{tab:test} gives the scores on the test seasons. Without
post-processing, the two U-Nets have the best value of every score except the
bias and the P99 ratio, with an RMSE of 8.58--8.61\,\mmd, a correlation of
0.613--0.621, an \FSS{} of 0.501--0.516 and a \CSI{40} of 0.162--0.180. DeepESD
comes next on most scores (correlation 0.581, \FSS{} 0.464, \CSI{40} 0.143).
ERA5 precipitation has a high RMSE (10.71\,\mmd) because it places heavy rain in
the right range of
intensities but often at the wrong place: 71\% of its predictions above
40\,\mmd{} are false alarms, against 53\% for the U-Net, and its \CSI{40} is
0.125. BCSD corrects its distribution and adds the observed fine-scale
climatology, which raises the correlation to 0.477, \FSS{} to 0.446 and
\CSI{40} to 0.141, but also the RMSE, to 11.18\,\mmd, the highest of the
baselines. One member of the GAN has the highest RMSE of all models
(11.71\,\mmd) and a correlation of 0.421, as expected for a sample of the rain
rather than an estimate of it (Section~\ref{sec:gan-results}). Most scores are
better on the test seasons than on the validation seasons (Table~\ref{tab:val}),
on which the U-Nets also have the best scores.

\begin{table}[t]
\caption{Scores on the test seasons (2021--2024, all pixel-days of the study
domain), without post-processing (top) and after quantile mapping (bottom).
RMSE, MAE and bias in \mmd. The best value of each column within each block is
in bold; bias and P99 ratio have no best value.}\label{tab:test}
\centering
\small
\setlength{\tabcolsep}{0pt}
\begin{tabular*}{\textwidth}{@{\extracolsep{\fill}}lccccccccc@{}}
\toprule
 & RMSE & MAE & Bias & $r$ & P99 & \FSS & \CSI{5} & \CSI{20} & \CSI{40} \\
 &      &     &      &     & ratio &     &         &          &          \\
\midrule
\multicolumn{10}{@{}l}{\emph{Raw outputs}}\\
Climatology & 9.94 & 4.82 & $-$0.13 & 0.386 & 0.34 & 0.064 & 0.334 & 0.025 & 0.000 \\
ERA5 precipitation & 10.71 & 4.17 & $-$0.38 & 0.465 & 0.85 & 0.410 & 0.399 & 0.188 & 0.125 \\
BCSD & 11.18 & 4.54 & $+$0.12 & 0.477 & 1.05 & 0.446 & 0.386 & 0.215 & 0.141 \\
DeepESD & 8.88 & 3.55 & $-$1.07 & 0.581 & 0.67 & 0.464 & 0.416 & 0.223 & 0.143 \\
U-Net & 8.61 & 3.43 & $-$0.85 & 0.613 & 0.71 & 0.501 & 0.455 & 0.254 & 0.162 \\
U-Net $+$ ERA5 precip. & \textbf{8.58} & \textbf{3.41} & $-$0.78 & \textbf{0.621} & 0.76 & \textbf{0.516} & \textbf{0.462} & \textbf{0.266} & \textbf{0.180} \\
Residual WGAN (one member) & 11.71 & 4.66 & $-$0.15 & 0.421 & 1.02 & 0.427 & 0.358 & 0.182 & 0.115 \\
\midrule
\multicolumn{10}{@{}l}{\emph{After quantile mapping}}\\
ERA5 precipitation & 10.83 & 4.40 & $-$0.22 & 0.480 & 1.00 & 0.441 & 0.377 & 0.211 & 0.139 \\
BCSD & 10.88 & 4.44 & $-$0.15 & 0.480 & 1.01 & 0.443 & 0.378 & 0.212 & 0.140 \\
DeepESD & 10.10 & 4.05 & $-$0.26 & 0.554 & 1.03 & 0.502 & 0.407 & 0.257 & 0.184 \\
U-Net & 9.96 & 3.98 & $+$0.00 & 0.578 & 1.06 & 0.515 & 0.439 & 0.273 & 0.186 \\
U-Net $+$ ERA5 precip. & \textbf{9.83} & \textbf{3.89} & $-$0.08 & \textbf{0.587} & 1.05 & \textbf{0.523} & \textbf{0.447} & \textbf{0.277} & \textbf{0.192} \\
Residual WGAN (one member) & 11.24 & 4.62 & $-$0.10 & 0.440 & 0.99 & 0.434 & 0.361 & 0.187 & 0.119 \\
\bottomrule
\end{tabular*}
\end{table}

The U-Net with ERA5 precipitation is better than the U-Net on every score of
Table~\ref{tab:test} except the bias and the P99 ratio, without post-processing
and after quantile mapping. The differences are small (0.008 in correlation,
0.015 in \FSS{} and 0.018 in \CSI{40} for the raw outputs).

The deterministic networks are too dry: the mean bias is $-0.78$ and
$-0.85$\,\mmd{} for the U-Nets and $-1.07$\,\mmd{} for DeepESD, against a mean
observed precipitation of 4.17\,\mmd{} over the test seasons, and their P99
ratio is 0.67--0.76. Quantile mapping reduces the mean bias of the U-Nets to
$-0.08$ and 0.00\,\mmd{} and brings their P99 ratio to 1.05--1.06. It
raises their \FSS{} by 0.006 to 0.014, their \CSI{20} by 0.011 to 0.019 and
their \CSI{40} by 0.012 to 0.024, while their RMSE increases by 1.25 to
1.35\,\mmd{} and their correlation decreases by 0.03. For the U-Net, the
POD at 40\,\mmd{} rises from 0.20 to 0.35, but the FAR also rises, from 0.53 to
0.72.

After quantile mapping, the U-Nets still have the best scores, but DeepESD comes
close to the U-Net on the event scores: its \FSS, 0.502, is 0.013 below that of
the U-Net, and its \CSI{40}, 0.184, is 0.002 below. The U-Net with ERA5
precipitation leads DeepESD more clearly, by 0.021 in \FSS{} and 0.008 in
\CSI{40}, and both U-Nets keep a clear lead in correlation (0.578 and 0.587,
against 0.554) and RMSE. The other baselines
remain well behind: ERA5 precipitation after quantile mapping and BCSD, with or
without it, reach an \FSS{} of 0.441 to 0.446 and a \CSI{40} of 0.139 to 0.141.
Since BCSD already corrects the distribution itself, a second mapping lowers its
\FSS{} and critical success indices by up to 0.008.

\subsection{Daily fields}

\begin{figure}[p]
\centering
\includegraphics[width=\textwidth,height=0.9\textheight,keepaspectratio]{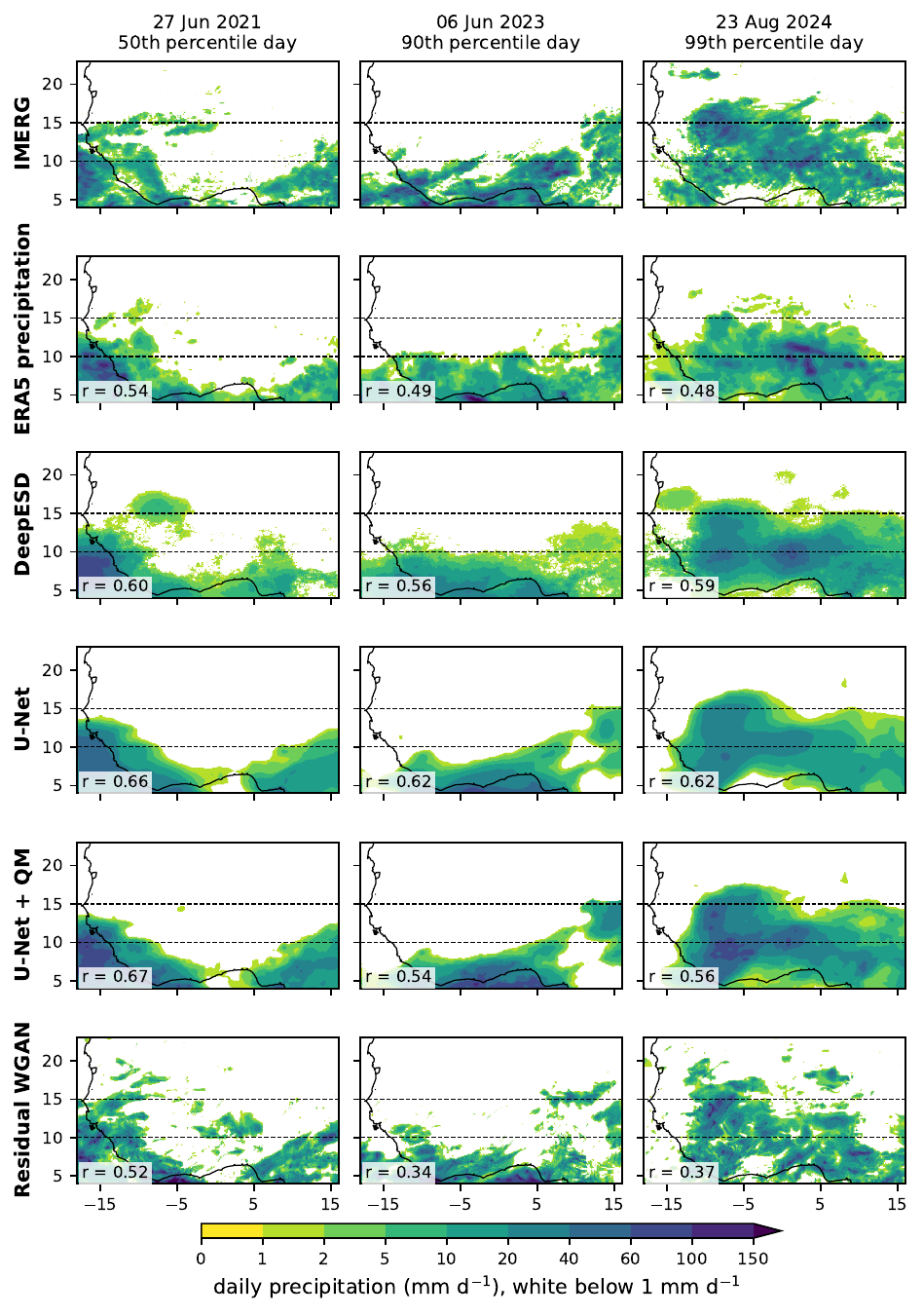}
\caption{Daily precipitation on three test days at the 50th, 90th and 99th
percentiles of the domain-mean observed precipitation (4.1, 6.4 and 8.2\,\mmd):
IMERG, ERA5 precipitation, DeepESD, the U-Net, the U-Net after quantile mapping,
and one member of the residual Wasserstein GAN. Values
below 1\,\mmd{} are white. The number in each panel is the spatial correlation
with IMERG on that day.}\label{fig:examples}
\end{figure}

Figure~\ref{fig:examples} shows three test days chosen objectively, at the 50th,
90th and 99th percentiles of the domain-mean observed precipitation. The U-Net
reproduces the large-scale envelope of the rain area, with spatial correlations
of 0.62 to 0.66 (DeepESD 0.56 to 0.60, ERA5 precipitation 0.48 to 0.54), but its
fields are much smoother than the observed ones and lack the individual
convective cells. Quantile mapping raises the intensities inside the predicted
rain area and changes its shape little. The GAN member has fine-scale structure
like that of IMERG, but not at the observed places, and its spatial correlation
drops to 0.34 to 0.52.

\subsection{Spatial structure of the errors}

\begin{figure}[p]
\centering
\includegraphics[width=\textwidth,height=0.88\textheight,keepaspectratio]{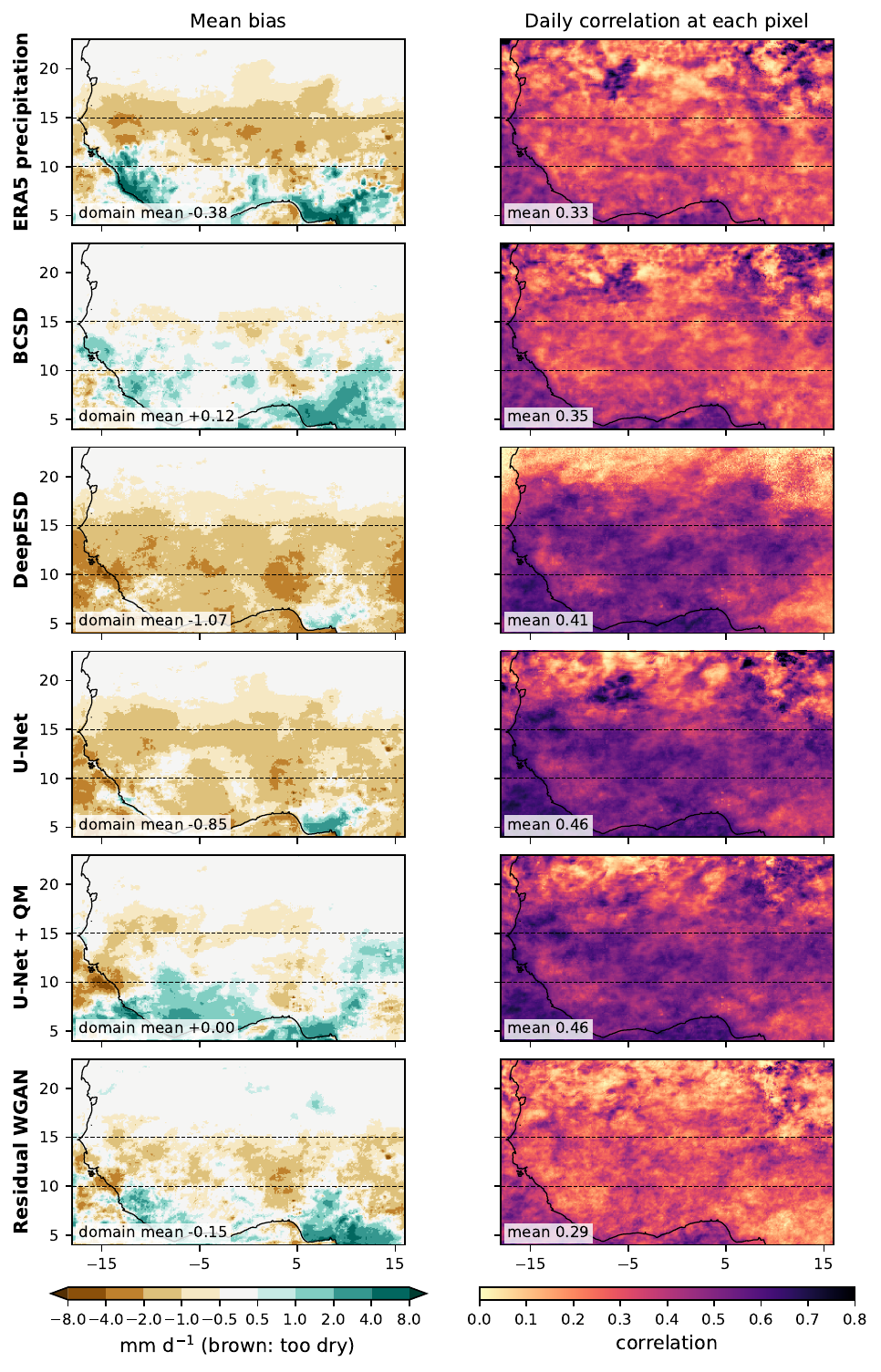}
\caption{Left: mean bias over the test seasons (\mmd; brown too dry, green too
wet), with its domain mean. Right: daily correlation with IMERG at each pixel,
with its domain mean. Rows: ERA5
precipitation, BCSD, DeepESD, the U-Net without and with quantile mapping, and one
member of the residual Wasserstein GAN; the U-Net with ERA5 precipitation is not
shown. The dashed lines separate the three zones.}\label{fig:maps}
\end{figure}

The dry bias of the raw U-Net covers most of the domain and is largest in the
Sudanian zone and along the western coast (Figure~\ref{fig:maps}, left); the
U-Net is too wet only in the south-east, near the coasts of Nigeria and
Cameroon. The dry bias of DeepESD has the same pattern and is stronger. Quantile
mapping removes the domain-mean bias of the U-Net but leaves regional biases of
1 to 4\,\mmd, locally more: too dry over Senegal, Guinea-Bissau and western
Guinea, too wet over the central Guinean zone and in the east. ERA5
precipitation is too wet along the Guinean coast and too dry inland. BCSD removes
most of these biases (domain mean $+0.12$\,\mmd) but is too wet by 1 to
4\,\mmd{} in the south-east and along the coast of Guinea. The GAN
member has a small domain-mean bias ($-0.15$\,\mmd) but strong regional
biases: too dry along the western coast and over much of the Sudanian zone, too
wet inland of the Guinean coast and in the south-east.

The daily correlation at each pixel (Figure~\ref{fig:maps}, right) averages 0.46
for the U-Net, with and without quantile mapping, against 0.41 for DeepESD, 0.35
for BCSD, 0.33 for ERA5 precipitation and 0.29 for one member of the GAN.

\subsection{Zones and months}

\begin{figure}[t]
\centering
\includegraphics[width=\textwidth]{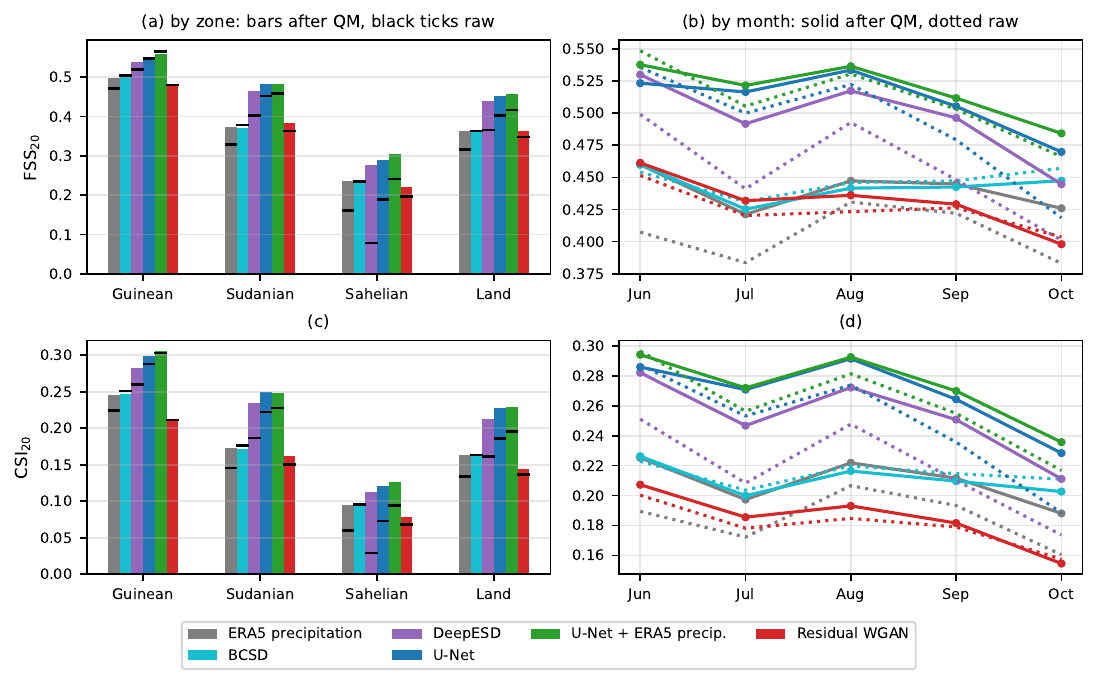}
\caption{\FSS{} (a, b) and \CSI{20} (c, d) on the test seasons, by zone and over
land (a, c) and by month (b, d). Bars (a, c) and solid lines (b, d) show the
scores after quantile mapping (QM); black ticks (a, c) and dotted lines (b, d) show
the raw scores. The GAN is represented by one member.}\label{fig:zones}
\end{figure}

Skill decreases from the Guinean to the Sahelian zone for every model
(Figure~\ref{fig:zones}). For the raw U-Net, \FSS{} is 0.547 in the Guinean
zone, 0.452 in the Sudanian zone and 0.189 in the Sahelian zone, and \CSI{20} is
0.288, 0.222 and 0.073. Quantile mapping helps most in the Sahel, where it
raises the \FSS{} of the U-Net from 0.189 to 0.289 and that of the U-Net with
ERA5 precipitation from 0.241 to 0.305; the latter has the highest correlation,
\FSS, \CSI{20} and \CSI{40} in the Sahel after quantile mapping, but the scores
there are low. Over land, the U-Nets reach an \FSS{} of 0.452--0.457 after
quantile mapping, against 0.440 for DeepESD, 0.363 for ERA5 precipitation and
0.362 for BCSD.

Without post-processing, a U-Net has the highest correlation and \FSS{} in every
zone and every month, and the highest \CSI{20} except in the Sahel, where BCSD,
with its own correction of the distribution, scores slightly higher (0.096
against 0.073 and 0.094). After quantile mapping, a U-Net has the highest
correlation, \FSS, \CSI{20} and \CSI{40} everywhere. Without the precipitation
input, the U-Net loses this lead in some zones and months, for instance to BCSD
in the Sahel and in October without post-processing (\FSS{} 0.235 against 0.189
and 0.457 against 0.419) and to DeepESD in June after quantile mapping (0.530
against 0.523). For the heaviest rain, the raw U-Nets have a lower \CSI{40} than
BCSD in the Sahel (0.014 and 0.031 against 0.040) and over land (0.056 and 0.066
against 0.079), where ERA5 precipitation also scores higher (0.068). The \FSS{}
and \CSI{20} of the U-Nets are lowest in October, at the end of the rainy
season.

\subsection{Spatial spectra and intensity distribution}\label{sec:spectra}

\begin{figure}[t]
\centering
\includegraphics[width=\textwidth]{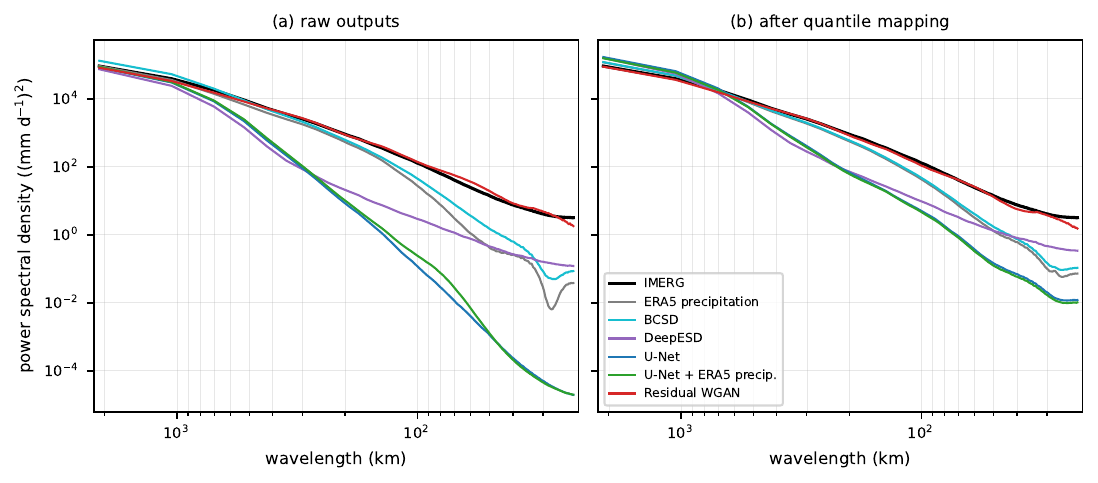}
\caption{Radially averaged power spectral density of the daily fields over the
test seasons, for IMERG and the models, without post-processing (a) and after
quantile mapping (b). Spectra are computed with pysteps on two $190\times190$
windows, each with its mean removed and a Hann taper, and averaged over the
windows and the days. The curve of the GAN is that of one member.}\label{fig:rapsd}
\end{figure}

Figure~\ref{fig:rapsd} compares the power spectra of the fields. ERA5
precipitation follows the observed spectrum down to wavelengths of a few hundred
kilometres (70\% of the observed power at 300\,km) but loses power at shorter
wavelengths, with 22\% of the observed power at 100\,km and 3\% at 50\,km (the
dip near 28\,km is the null of the bilinear interpolation from the
0.25\textdegree{} grid). BCSD, which multiplies the interpolated ratios by the
observed fine-scale climatology, keeps more of the small-scale power (83, 35 and
11\% of the observed power at 300, 100 and 50\,km) but has 1.36 times the
observed power at 1000\,km. The trained deterministic networks lose power at
much larger scales: without post-processing, the U-Nets and DeepESD have
3--4\% of the observed power at 300\,km, and at 100\,km the U-Nets have less
than 0.2\%, against 2\% for DeepESD.
Quantile mapping does not restore the small scales: after mapping, the U-Nets
have 15--16\% of the observed power at 300\,km and 4\% at 100\,km, while
their power at the largest scales, around 1000\,km, rises to 1.5--1.7 times the
observed one. A member of the GAN, in contrast, has nearly the observed power at
all scales (86, 105, 109 and 85\% of it at 1000, 300, 100 and 25\,km), with an
excess of 1.37 times the observed power near 50\,km.

The distribution of intensities leads to similar conclusions. IMERG exceeds
1\,\mmd{} on 33.0\% of the pixel-days, 20\,\mmd{} on 6.2\% and 50\,\mmd{} on
1.1\%. The raw U-Net has about the right frequency of rain above 1\,\mmd{}
(32.6\%) but exceeds 20\,\mmd{} on only 4.1\% of the pixel-days and 50\,\mmd{}
on 0.4\%, and reaches at most 107\,\mmd. ERA5 precipitation rains too often
(38.5\% above 1\,\mmd) and too rarely above 20\,\mmd{} (3.9\%), whereas BCSD
is close to the observed frequencies (31.9, 6.8 and 1.2\% above 1, 20 and
50\,\mmd). After quantile
mapping, the U-Net exceeds 20 and 50\,\mmd{} on 6.6\% and 1.3\% of the
pixel-days, close to the observations; the raw GAN member exceeds them on 5.9\%
and 1.1\%.

Above the 99th percentile, the mapped distributions fall short of the observed
one: the 99.9th percentile of the pixel-day values is 90 to 96\% of the observed
one for every model of Table~\ref{tab:test} (U-Net 93\%), and amounts above
100\,\mmd{} are 66 to 88\% as frequent as in IMERG (U-Net 74\%). Without mapping, the U-Nets almost never exceed
100\,\mmd{} (0.2\% of the observed frequency for the U-Net), whereas ERA5
precipitation and a GAN member both exceed it 1.4 times as often as IMERG.
BCSD, whose own mapping extends to the largest values of its training samples,
has a 99.9th percentile 97\% of the observed one and exceeds 100\,\mmd{} 0.91
times as often as IMERG.

\subsection{The GAN ensemble}\label{sec:gan-results}

\begin{figure}[p]
\centering
\includegraphics[width=\textwidth]{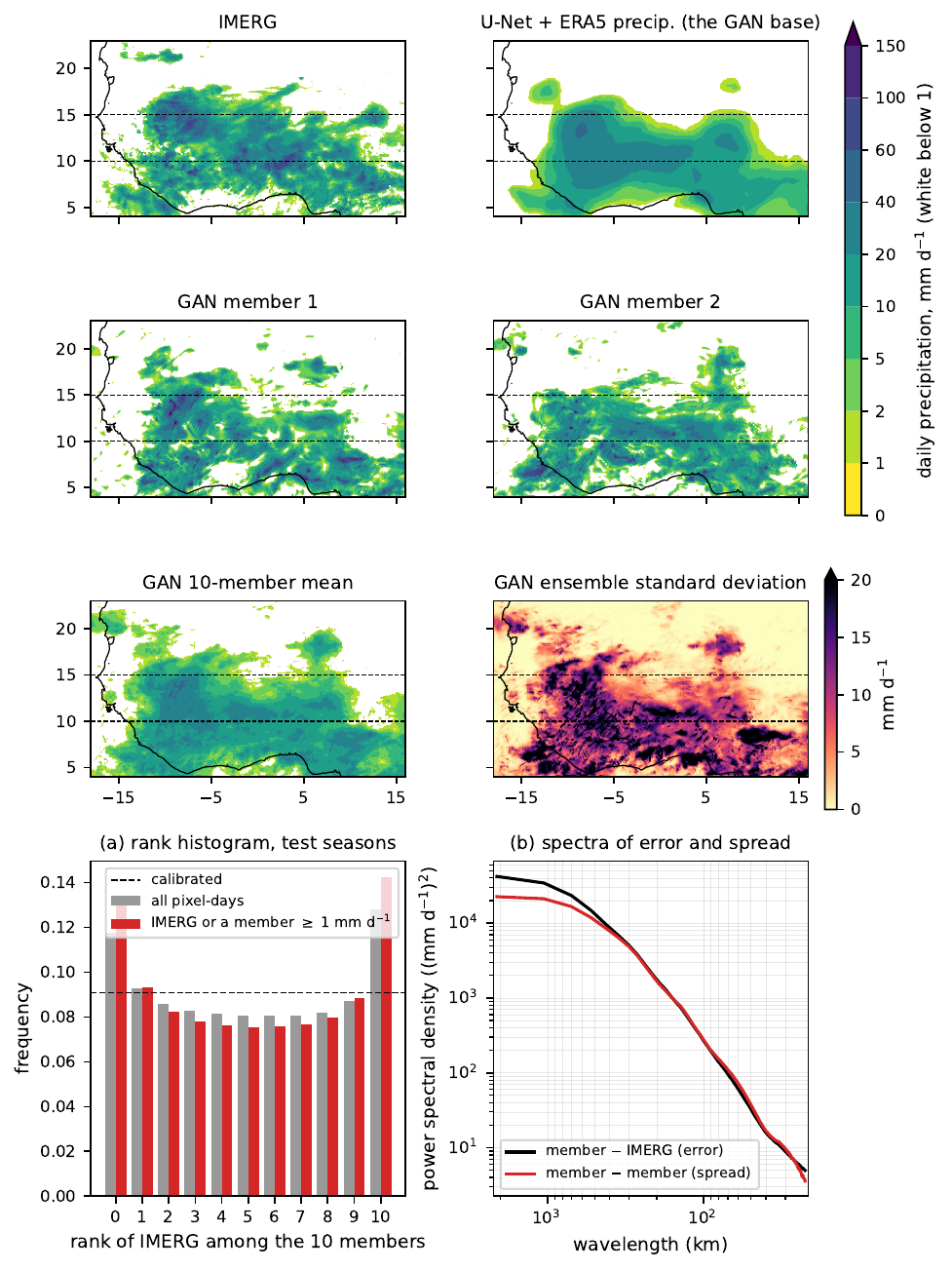}
\caption{Top: the test day at the 99th percentile of domain-mean precipitation
(23~August 2024): IMERG, the prediction of the U-Net with ERA5 precipitation, on
which the GAN builds, two members of the GAN, and the mean and the standard
deviation of its ten members.
(a) Rank of IMERG among the ten members over all test pixel-days (grey) and over
the pixel-days on which IMERG or at least one member has at least 1\,\mmd{}
(red); amounts below 0.1\,\mmd{} count as zero and ties are broken at random;
the dashed line is the flat histogram of a calibrated ensemble. (b) Power
spectra, computed as in Figure~\ref{fig:rapsd}, of the difference between a
member and IMERG (error, five members) and between two members (spread, five
pairs), averaged over the test days; for a calibrated ensemble the two curves
coincide.}\label{fig:gan}
\end{figure}

Table~\ref{tab:gan} gives the scores of the ensemble of the GAN. The ten-member
ensemble is close to calibrated at the pixel scale: its spread-skill ratio is
0.92 on the test seasons, and its CRPS, 2.56\,\mmd, is 25\% lower than the MAE
of the U-Net with ERA5 precipitation it builds on, 3.41\,\mmd, which is the CRPS
of that network. It is under-dispersive at the largest scales: the spread-skill
ratio of fields averaged over 2\textdegree{} boxes is 0.76, and the power of the
differences between members is 62\% of that of the errors at 1000\,km, against
92 to 115\% at 300, 100, 50 and 25\,km (Figure~\ref{fig:gan}b). On the
pixel-days on which IMERG or at least one member has at least 1\,\mmd, IMERG
lies outside the range of the members 27\% of the time (14\% above and 13\%
below), against 18\% for a calibrated ensemble (Figure~\ref{fig:gan}a). The
members have nearly the observed power spectrum (Section~\ref{sec:spectra}), and
the rain above 20\,\mmd{} forms 96 separate
areas per day, with a median size of 3 cells, against 101 areas of 3 cells in
IMERG and 2 areas of 514 cells for the U-Net with ERA5 precipitation. The RMSE
of the ensemble mean, 8.99\,\mmd, is higher than that of this U-Net,
8.58\,\mmd, partly because of the sampling error of a mean of ten members, and
its \FSS{} is 0.494, against 0.516.

\begin{table}[t]
\caption{The ensemble of the GAN on the test seasons (ten members). Top: CRPS (\mmd), spread-skill ratio
(SSR) at the pixel scale and for 2\textdegree{} boxes, percentage of the pixel-days with IMERG outside the range of the
members, among those on which IMERG or a member has at least 1\,\mmd{} (18\% for a calibrated ensemble), and RMSE
(\mmd) and \FSS{} of the ensemble mean; for the U-Net with ERA5 precipitation, on which the GAN builds, the CRPS is
the MAE and the ensemble mean is the prediction. Bottom: ratio of the power of the differences between members to
that of the errors (s/e) at 300, 100 and 50\,km, ratio of the power of one member to that of IMERG ($P$) at 100 and
50\,km, and number per day and median size (cells) of the areas with at least 20\,\mmd.}\label{tab:gan}
\centering
\footnotesize
\setlength{\tabcolsep}{0pt}
\begin{tabular*}{\textwidth}{@{\extracolsep{\fill}}lcccccc@{}}
\toprule
 & CRPS & SSR & SSR 2\textdegree & Outside (\%) & RMSE mean & \FSS{} mean \\
\midrule
GAN & 2.56 & 0.92 & 0.76 & 27 & 8.99 & 0.494 \\
U-Net $+$ ERA5 precip. & 3.41 & --- & --- & --- & 8.58 & 0.516 \\
\midrule
 & s/e 300\,km & s/e 100\,km & s/e 50\,km & $P$ 100\,km & $P$ 50\,km & Areas (size) \\
\midrule
GAN & 0.96 & 1.03 & 1.15 & 1.09 & 1.37 & 96 (3) \\
U-Net $+$ ERA5 precip. & --- & --- & --- & $<$0.01 & $<$0.01 & 2 (514) \\
IMERG & --- & --- & --- & 1 & 1 & 101 (3) \\
\bottomrule
\end{tabular*}
\end{table}

\subsection{Predictor importance}

\begin{figure}[t]
\centering
\includegraphics[width=0.62\textwidth]{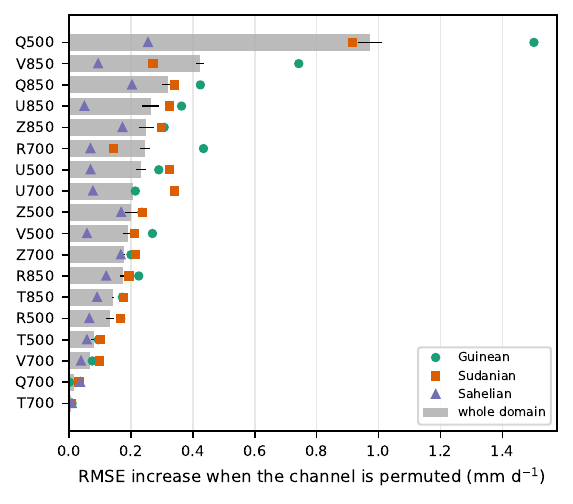}
\caption{Increase in the RMSE of the U-Net on the test seasons when one input
channel is permuted between days, for the whole domain (bars, with the standard
deviation over three permutations) and for each zone (symbols). The RMSE without
permutation is 8.61\,\mmd.}\label{fig:permutation}
\end{figure}

Permuting the 500\,hPa specific humidity increases the RMSE of the U-Net by
0.97\,\mmd, more than twice as much as any other predictor
(Figure~\ref{fig:permutation}). It is followed by meridional wind at 850\,hPa
(0.42\,\mmd), specific humidity at 850\,hPa (0.32\,\mmd) and zonal wind at
850\,hPa (0.26\,\mmd), while temperature at 700\,hPa has almost no effect
(0.01\,\mmd). The dominance of mid-level humidity comes from the Guinean and
Sudanian zones (1.50 and 0.91\,\mmd); in the Sahel, where the errors are small,
no predictor stands out and the increases are at most 0.26\,\mmd.

\section{Discussion}\label{sec:discussion}

\subsection{What the U-Nets and the precipitation input add}

Over the whole domain, the U-Nets have the best scores of all models, raw and
after quantile mapping. Against ERA5 precipitation, which contains the ERA5
model's own estimate of rainfall, the U-Net gains 0.15 in correlation, 0.09 in
\FSS{} and 0.04 in \CSI{40}; once both are quantile-mapped, the gains are 0.10,
0.07 and 0.05. The coarse precipitation field is thus a strong baseline for the detection
of heavy rain, and the natural reference for a downscaling model in this region.
Its classical statistical downscaling does not change this picture: BCSD scores
about as well as mapped ERA5 precipitation, and the mapped U-Net leads it by 0.10
in correlation, 0.07 in \FSS{} and 0.05 in \CSI{40}. Fitted on the training
seasons, BCSD is too wet on the later seasons ($+0.23$\,\mmd{} on the validation
and $+0.12$\,\mmd{} on the test seasons). The domain-mean ERA5 precipitation is
89\% of that of IMERG over the training seasons but 92\% over the validation and
91\% over the test seasons; it ranges from 84 to 98\% between seasons, with the
highest values from 2015 to 2019. A correction fitted on past seasons of the
reanalysis carries such changes over to the corrected values.

DeepESD, whose single dense layer connects the coarse predictors of the whole
domain to every pixel, is the closest competitor, however, it sees the predictors averaged over 1\textdegree{}
boxes (Section~\ref{sec:baselines}), so its gap to the U-Nets combines the
effect of the architecture with that of the coarser input.

Adding ERA5 precipitation to the U-Net gives small gains, which
make it the best model, without post-processing and after quantile mapping. The
gains are largest in the Sahel, where the U-Net is least skilful (\FSS{} 0.241
against 0.189 without post-processing; Figure~\ref{fig:zones}), and at most 0.02 in correlation, \FSS{}
and the critical success indices over the whole domain, presumably because much
of the information in the precipitation field is already in the humidity and
circulation predictors from which the ERA5 model computes it. 
A model that needs this input, however, is no longer a perfect-prognosis model:
applied to the output of a climate model, it would take in the simulated
precipitation and its biases. For such applications, the U-Net without this
input is preferable, at a small cost in skill.

\subsection{The loss function, quantile mapping and the missing small scales}\label{sec:disc-loss}

The dry bias of the deterministic networks (Figure~\ref{fig:maps}) follows from the loss. The
absolute-error term of Equation~\eqref{eq:asym} is minimised by the conditional
median, which for daily rainfall, zero-inflated and right-skewed, lies well
below the conditional mean; the second term raises the predictions only where
the observed rain is heavy for its location. The resulting fields have a
realistic frequency of light rain (Section~\ref{sec:spectra}) but too little
heavy rain and a mean deficit of about a fifth to a quarter of the observed
mean. A loss of this form was designed to reproduce the whole distribution, with
a penalty on underestimation that grows with the rarity of the observed
value~\citep{doury2024}. \citet{gonzalezabad2025} found it the best of the
deterministic losses they compared for heavy precipitation over Spain, while
still underestimating the largest extremes; in our setting too, it does not
remove the need for a correction of the distribution.

Quantile mapping provides that correction at a known cost. It restores the mean
and the upper quantiles and improves the detection of heavy rain, but it
inflates the variance of a prediction that is only partly correlated with the
observations and therefore increases the RMSE. \citet{vonstorch1999} and, for
quantile mapping, \citet{maraun2013} discuss the limits of such an inflation of
the variance in downscaling. The spectra (Figure~\ref{fig:rapsd}) show where this variance goes. The
mapping is a pointwise transformation of a smooth field (Figure~\ref{fig:examples}), so it amplifies the
large scales beyond their observed power and raises the small scales only to a
few percent of their observed power at 100\,km. A deterministic prediction at
this resolution is an estimate of the rain that the large-scale state makes
likely, not a realisation of it, and the missing small-scale variance is the
part of the rainfall that the predictors do not determine. \citet{lee2026} found
the same loss of high-frequency power for deterministic networks over
sub-Saharan Africa, even with the coarse precipitation field among the inputs.

\subsection{Calibration of the GAN ensemble}\label{sec:disc-ens}

The GAN shows that neither the ensemble-mean squared error nor a critic that
scores one member at a time prevents a calibrated ensemble: its spread-skill
ratio is 0.92. The ensemble-mean squared error still penalises the
spread, by its variance divided by the number of members, rather than rewarding
it, so this calibration is a property of the trained model and not a guarantee
of the loss. The slight under-dispersion that remains, 27\% of the observations
outside the range of the members and a spread-skill ratio of 0.76 for
2\textdegree{} boxes (Table~\ref{tab:gan} and Figure~\ref{fig:gan}a), is consistent with the intercomparison of
\citet{rampal2025intercomparison}, who found ResGAN-v2 slightly
under-dispersive.

The small scales of daily rainfall at 0.1\textdegree{} are largely not
determined by the ERA5 predictors (Section~\ref{sec:disc-loss}), so an ensemble
reproduces them only if its members generate them, as those of the GAN do, with
nearly the observed spectrum and number of heavy-rain areas. Diffusion and
flow-matching models, which were better calibrated than GANs in the
intercomparison of \citet{rampal2025intercomparison} and restore the observed
spectrum when conditioned on coarse precipitation~\citep{lee2026}, are an
alternative; in these two studies, they were trained on regional climate-model
output or conditioned on a coarse-grained version of the target, rather than on
reanalysis predictors against observations. The corrective diffusion model of
\citet{mardani2025}, which models the residual of a regression U-Net, is the
direct counterpart of our residual GAN, but its sampling takes tens of network
evaluations per member instead of one, which would multiply the cost of
generating an ensemble (Appendix~\ref{app:timing}).

\subsection{Predictors}

The importance of mid-tropospheric humidity (Figure~\ref{fig:permutation}) is consistent with the sensitivity
of tropical deep convection to the moisture of the free
troposphere~\citep{holloway2009}, and that of the low-level meridional wind
with its role in carrying the monsoon flow inland~\citep{nicholson2013}. Permutation importance should nevertheless be read
with care: specific and relative humidity at the same level are strongly
correlated, permuting one of them creates combinations of inputs that never
occur~\citep{hooker2021}, and the importance of a predictor is partly shared
with its correlated neighbours~\citep{mcgovern2019}. The weak importance of all
predictors in the Sahel probably reflects the low skill of all models there
(Figure~\ref{fig:zones}) rather than a different physics: most of the rainfall there comes from a small
number of organised convective systems~\citep{mathon2002}, many of them
associated with African easterly waves~\citep{fink2003,kiladis2006}, and the
daily large-scale fields may constrain the position of these systems only
loosely.

\subsection{Limitations}

The target is a satellite product, not a gauge analysis, and IMERG has its own
errors over West Africa. For earlier versions of the product, the agreement with
gauges was weaker in the Sahel than in the humid south~\citep{dezfuli2017}, and
in the forest zone IMERG overestimated rain from weak convection and
underestimated rain from strong mesoscale convective
systems~\citep{maranan2020}; the models are trained to reproduce IMERG, errors
included. The
training set has 2\,907 days, which limits the complexity that can be learnt:
the U-Nets reach their best validation loss between epochs 23 and 36 of 40 and
overfit afterwards (Figure~\ref{fig:training}).

Rainfall in this region is also hard to predict: global ensemble forecasts of 1-
to 5-day rainfall totals over northern tropical Africa hardly beat climatology,
even after statistical post-processing, a failure that \citet{vogel2018}
attribute partly to the parameterisation of convection. Our models are given the
analysed large-scale state rather than a forecast of it, but their skill remains
limited: the correlation between the predicted and observed daily series at a
pixel averages 0.46 for the U-Net (Figure~\ref{fig:maps}).

\section{Conclusions}\label{sec:conclusion}

We downscaled daily ERA5 atmospheric fields to IMERG~V07 precipitation at
0.1\textdegree{} over West Africa for the rainy seasons of 2000--2024, with all
choices based on two validation seasons and four test seasons kept for the
evaluation. U-Nets trained with an asymmetric loss are the most skilful models
over the domain, ahead of DeepESD and of the simpler baselines, including the
bias correction and spatial disaggregation of ERA5 precipitation; DeepESD comes
close to the perfect-prognosis U-Net on the event scores once both are
quantile-mapped. ERA5 precipitation as an extra input gives small gains, at the price of leaving the perfect-prognosis
formulation.

The deterministic networks are too dry, underestimate heavy rain and lack almost
all of the observed variance at scales below a few hundred kilometres. A
quantile mapping fitted on the validation seasons corrects the distribution and
improves the detection of heavy rain, at the cost of a larger RMSE, but it does
not restore the small scales, and the upper tail of the mapping must be treated
with care. A residual Wasserstein GAN built on the U-Net with ERA5
precipitation and trained with the losses of ResGAN-v2 gives a ten-member
ensemble that is close to calibrated at the pixel scale (spread-skill ratio 0.92)
and slightly under-dispersive at the largest scales. Its members have nearly the
observed power spectrum and number of heavy-rain areas, and its CRPS,
2.56\,\mmd, is 25\% below the MAE of the network it builds on. Mid-tropospheric
humidity is the dominant predictor of the U-Net outside the Sahel.

Once trained, a U-Net downscales one day in 3 to 10\,ms on a GPU and in less
than a second on a desktop processor without a GPU (Table~\ref{tab:timing}).
Training is the costly step: about 16 minutes on a recent GPU, against an
estimated 34 hours on eight threads of a desktop processor.

Applying the models to climate projections, verifying them against rain gauges
and comparing the GAN with diffusion models are the natural next steps.

\section*{Acknowledgements}

The computations were carried out on the ESPRI facility of the IPSL Data
and Computing Center, supported by CNRS, Sorbonne Universit\'e, CNES and
\'Ecole Polytechnique, from which the ERA5 and IMERG data were also accessed.
D.~Sagna acknowledges LOSEC, Universit\'e Assane Seck de Ziguinchor, and the IRD
through the LMI ECLAIRS2 and the JEAI-CLIMAS programme.

\textit{Use of generative AI.} The generative AI model Claude Opus~5.5 (Anthropic) assisted the authors at
different stages of the work (code, figures and text); the authors checked all results and the text and take
full responsibility for the content of this article.

\textit{Funding.} R.~Lguensat received funding managed by the Agence Nationale de la
Recherche under France 2030 bearing the reference ANR-22-EXTR-0011
(TRACCS-PC10-LOCALISING project).

\section*{Data and code availability}

IMERG~V07 is available from NASA GES DISC~\citep{imergv07}, ERA5 from the
Copernicus Climate Change Service~\citep{era5pl,era5sl} and ETOPO~2022 from NOAA
NCEI~\citep{etopo2022}. This study contains modified Copernicus Climate Change
Service information (2026); neither the European Commission nor ECMWF is
responsible for any use that may be made of the Copernicus information or data
it contains. The daily datasets built for this study, the model
outputs and the trained weights are openly available on Zenodo at
\url{https://doi.org/10.5281/zenodo.23188297}~\citep{sagna2026zenodo}.

The code for data preparation, training, evaluation, figures and execution
times is openly available, under the MIT licence, in the same Zenodo
record~\citep{sagna2026zenodo}.

\bibliographystyle{plainnat}
\bibliography{ref}

\appendix

\section{Choice of the quantile-mapping design}\label{app:qm}

The quantile mapping of Section~\ref{sec:qm} was chosen on the ERA5
precipitation baseline, which has no training data, by a two-fold
cross-validation within the validation seasons: the mapping is fitted on 2019
and applied to 2020, and conversely, and the scores are computed on the
concatenated predictions (Table~\ref{tab:qmcv}). We compared the treatment of
the values above the highest fitted quantile, with a monthly $5\times5$-pixel
pooling: extending the correction with a constant ratio or a constant
difference, or setting such values to the highest observed quantile, with
quantiles fitted up to the maximum or up to the 99th percentile. We also
compared the pooling, with the capped mapping fitted up to the 99th percentile.

Because the RMSE favours no mapping at all (10.72\,\mmd{} for the raw
predictions in the same cross-validation), we retained a variant whose mean bias
and P99 ratio are among the closest to the observations and whose event scores
are among the best: pooling over $5\times5$ pixels and the neighbouring months.
Pooling over $3\times3$ or $9\times9$ pixels with the neighbouring months gave
almost the same scores, and pooling over two months on each side improved \FSS{}
slightly at the cost of a P99 ratio of 0.96. Without spatial pooling, \FSS{} is
higher (0.410), but the mean bias and the P99 ratio move away from the
observations ($-0.04$ and 0.97) and \CSI{40} is lower; the U-Net shows the same
trade-off in this cross-validation (\FSS{} 0.494 against 0.486, \CSI{40} 0.161
against 0.165).

In the same cross-validation, we also ran the multiplicative quantile delta
mapping of \citet{cannon2015} and CDF-t~\citep{michelangeli2009} with
SBCK~\citep{robin2019}. We used the singularity stochastic removal of
\citet{vrac2016}, a limit on the upper tail, the same pooling and, for CDF-t, no
normalisation of the data (the default normalisation by the ratio of the
standard deviations inflates the nearly dry predictions of the U-Net north of
the rain belt). They give \FSS{} 0.397 and \CSI{40} 0.110--0.115 for the ERA5
baseline, against 0.404 and 0.119, and \FSS{} 0.482--0.486 and \CSI{40}
0.162--0.163 for the U-Net, against 0.486 and 0.165, with a larger RMSE in every
case. Both methods carry the change of the predicted distribution between the
calibration and application periods over to the corrected values, which is their
purpose for climate projections; between seasons of the same climate, that
change is sampling variability.

\begin{table}[t]
\caption{Two-fold cross-validation of quantile-mapping designs within the
validation seasons, for the ERA5 precipitation baseline. ``Ratio'' and
``difference'' extend the correction above the highest fitted quantile with a
constant ratio or difference; ``cap'' sets those values to the highest fitted
observed quantile. The retained design is in bold.}\label{tab:qmcv}
\centering
\small
\setlength{\tabcolsep}{0pt}
\begin{tabular*}{\textwidth}{@{\extracolsep{\fill}}lllccccccc@{}}
\toprule
Upper tail & Highest level & Pooling & RMSE & Bias & $r$ & P99 & \FSS & \CSI{20} & \CSI{40} \\
\midrule
\multicolumn{3}{@{}l}{No mapping} & 10.72 & $-$0.35 & 0.449 & 0.81 & 0.370 & 0.162 & 0.110 \\
Ratio      & maximum & month, $5\times5$ & 19.19 & $+$1.38 & 0.317 & 1.53 & 0.386 & 0.175 & 0.105 \\
Ratio      & 0.99    & month, $5\times5$ & 17.54 & $+$1.11 & 0.331 & 1.39 & 0.387 & 0.176 & 0.106 \\
Difference & maximum & month, $5\times5$ & 13.97 & $+$0.82 & 0.396 & 1.32 & 0.390 & 0.177 & 0.110 \\
Difference & 0.99    & month, $5\times5$ & 12.90 & $+$0.55 & 0.413 & 1.18 & 0.393 & 0.179 & 0.115 \\
Cap        & maximum & month, $5\times5$ & 12.87 & $+$0.59 & 0.406 & 1.21 & 0.391 & 0.178 & 0.111 \\
Cap        & 0.99    & month, $5\times5$ & 11.70 & $+$0.24 & 0.424 & 1.03 & 0.395 & 0.179 & 0.114 \\
Cap        & 0.99    & month, $3\times3$ & 11.74 & $+$0.25 & 0.421 & 1.03 & 0.397 & 0.179 & 0.112 \\
Cap        & 0.99    & month, $7\times7$ & 11.63 & $+$0.22 & 0.427 & 1.03 & 0.394 & 0.180 & 0.115 \\
Cap        & 0.99    & month, $9\times9$ & 11.57 & $+$0.21 & 0.429 & 1.02 & 0.394 & 0.180 & 0.116 \\
Cap        & 0.99    & $\pm$1 month, $1\times1$ & 11.10 & $-$0.04 & 0.442 & 0.97 & 0.410 & 0.184 & 0.118 \\
Cap        & 0.99    & $\pm$1 month, $3\times3$ & 11.24 & $+$0.00 & 0.441 & 0.99 & 0.405 & 0.184 & 0.119 \\
\textbf{Cap} & \textbf{0.99} & \textbf{$\pm$1 month, $5\times5$} & \textbf{11.18} & $\mathbf{-0.01}$ & \textbf{0.443} & \textbf{0.99} & \textbf{0.404} & \textbf{0.185} & \textbf{0.119} \\
Cap        & 0.99    & $\pm$1 month, $9\times9$ & 11.10 & $-$0.04 & 0.446 & 0.98 & 0.404 & 0.185 & 0.121 \\
Cap        & 0.99    & $\pm$2 months, $5\times5$ & 10.96 & $-$0.10 & 0.450 & 0.96 & 0.409 & 0.188 & 0.120 \\
\bottomrule
\end{tabular*}
\end{table}

\section{Training curves}\label{app:training}

\begin{figure}[t]
\centering
\includegraphics[width=\textwidth]{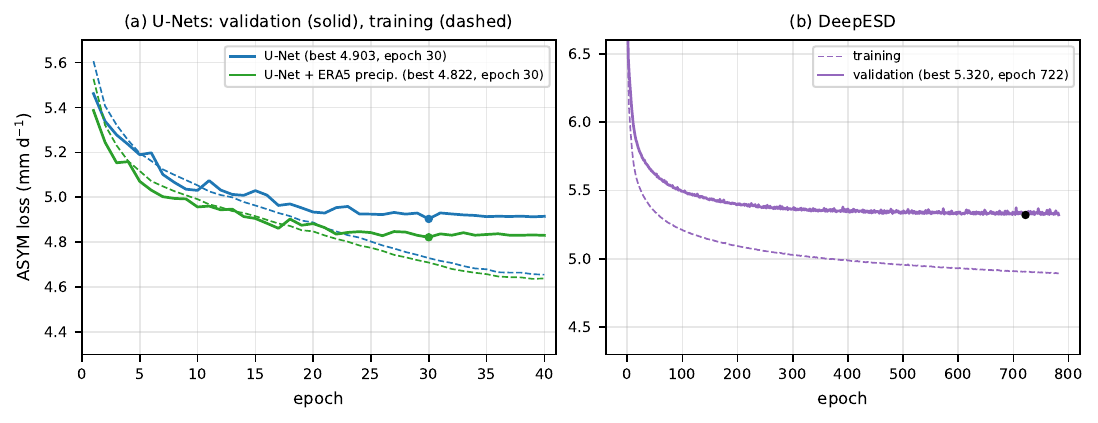}
\caption{Training (dashed) and validation (solid) loss of the U-Nets (a) and of
DeepESD (b). Dots mark the retained epochs; epochs are counted from
1.}\label{fig:training}
\end{figure}

Figure~\ref{fig:training} shows the training curves. Both
U-Nets reach their lowest validation loss at epoch 30, 4.90 for the U-Net and
4.82 with ERA5 precipitation, and overfit afterwards, while their training loss
keeps decreasing. DeepESD reaches its lowest validation loss, 5.32, at epoch 722.
Table~\ref{tab:val} gives the validation scores of the final models.

\begin{table}[t]
\caption{Scores on the validation seasons (2019--2020), without
post-processing. Same layout as Table~\ref{tab:test}.}\label{tab:val}
\centering
\small
\setlength{\tabcolsep}{0pt}
\begin{tabular*}{\textwidth}{@{\extracolsep{\fill}}lccccccccc@{}}
\toprule
 & RMSE & MAE & Bias & $r$ & P99 & \FSS & \CSI{5} & \CSI{20} & \CSI{40} \\
 &      &     &      &     & ratio &     &         &          &          \\
\midrule
Climatology & 9.94 & 4.87 & $-$0.19 & 0.383 & 0.34 & 0.058 & 0.340 & 0.022 & 0.000 \\
ERA5 precipitation & 10.72 & 4.32 & $-$0.35 & 0.449 & 0.81 & 0.370 & 0.396 & 0.162 & 0.110 \\
BCSD & 11.21 & 4.73 & $+$0.23 & 0.459 & 1.01 & 0.417 & 0.378 & 0.194 & 0.126 \\
DeepESD & 8.92 & 3.66 & $-$1.03 & 0.572 & 0.63 & 0.427 & 0.420 & 0.198 & 0.123 \\
U-Net & 8.67 & 3.51 & $-$0.95 & 0.602 & 0.65 & 0.459 & 0.452 & 0.222 & 0.140 \\
U-Net $+$ ERA5 precip. & \textbf{8.62} & \textbf{3.49} & $-$0.83 & \textbf{0.611} & 0.70 & \textbf{0.491} & \textbf{0.462} & \textbf{0.244} & \textbf{0.152} \\
Residual WGAN (one member) & 11.70 & 4.78 & $-$0.13 & 0.409 & 1.00 & 0.408 & 0.357 & 0.170 & 0.104 \\
\bottomrule
\end{tabular*}
\end{table}

\section{Execution time}\label{app:timing}

Table~\ref{tab:timing} gives the time needed to downscale one day with the
trained models on a recent high-end GPU (NVIDIA GeForce RTX~5090, 32\,GB,
released in 2025), on an older consumer GPU (GeForce RTX~2080~Ti, 11\,GB, 2018)
and on a desktop processor without a GPU (Intel Core i5-14600KF, 2023), with
PyTorch~2.10 in single precision (with TensorFloat-32 matrix products on the
RTX~5090, as in training). The time runs from the normalised coarse predictors
of the day, held in memory, to the precipitation field on the study domain: it
includes the interpolation of the predictors to the 0.1\textdegree{} grid, on a
GPU their copy to the GPU, and for DeepESD their averaging over 1\textdegree{}
boxes; it excludes reading the data from disk. The times are averaged over the
612 test days on the GPUs, with one day at a time and with batches of 16 days,
and over 24 test days on the processor, one day at a time, after two warm-up
batches.

A U-Net downscales one day in about 3\,ms on the RTX~5090 and 10\,ms on the
RTX~2080~Ti, with at most 0.3\,GB of GPU memory, and in 0.3 to 0.9\,s on the
processor, so a season of 153 days takes 0.4\,s on the RTX~5090, 1.5\,s on the
RTX~2080~Ti and at most 2.3 minutes on one processor thread. The ten members of
the GAN take about four to seven times as long as one prediction of the U-Net it
builds on, and DeepESD is one to two orders of magnitude faster than the U-Nets.
Quantile mapping adds about 4\,ms per day once fitted, and fitting it on the 306
validation days takes about one minute on the processor. BCSD, which trains no
network, takes about one minute for the 3\,825 days of 2000--2024 on eight
processor cores.

Training is the costly step. On one RTX~5090, a U-Net trains in about 16 minutes
(40 epochs), DeepESD in 2 minutes (782 epochs, with early stopping) and the GAN
in 2.2 hours (200 epochs). Without a GPU, one training step of the U-Net (a batch
of eight days) takes 8.4\,s on eight threads of the i5-14600KF, so a full
training would take about 34 hours, and 103 hours on one thread. A GPU is thus
needed to train the networks in a reasonable time, but not to apply them.

\begin{table}[t]
\caption{Time to downscale one day of the test seasons with the trained models
(ms per day), one day at a time and, on the GPUs, in batches of 16 days (in
brackets). The GAN generates ten members. CPU: Intel Core i5-14600KF, with
eight threads and with one thread.}\label{tab:timing}
\centering
\small
\setlength{\tabcolsep}{0pt}
\begin{tabular*}{\textwidth}{@{\extracolsep{\fill}}lcccc@{}}
\toprule
 & RTX~5090 & RTX~2080~Ti & CPU, 8 threads & CPU, 1 thread \\
\midrule
DeepESD & 0.18 (0.05) & 0.64 (0.39) & 2.6 & 4.3 \\
U-Net & 2.8 (2.2) & 9.7 (6.9) & 263 & 913 \\
U-Net $+$ ERA5 precip. & 2.9 (2.2) & 9.8 (6.9) & 295 & 860 \\
GAN, ten members & 20.0 (11.3) & 50.6 (34.3) & 1200 & 3340 \\
\bottomrule
\end{tabular*}
\end{table}

\end{document}